\documentstyle[prl,aps,epsf,epsfig,array]{revtex}

\newcommand{\be}{\begin{equation}}
\newcommand{\ee}{\end{equation}}
\newcommand{\bqn}{\begin{eqnarray}}
\newcommand{\eqn}{\end{eqnarray}}

\begin{document}
\title{Dressed-state approach to a simple radiating atom}

\author{G. Flores-Hidalgo$^{(a)}$\thanks{E-mail: gflores@ift.unesp.br}
and A. P. C. Malbouisson$^{(b)}$\thanks{E-mail: adolfo@cbpf.br}}

\address{$^{(a)}$ Instituto de F\'{\i}sica Teorica, UNESP, Rua Pamplona 145,
01405-900, S\~ao Paulo, SP, Brazil\\
$^{(b)}$ CBPF/MCT, Rua Dr. Xavier Sigaud 150, Urca, 22290-180, Rio
de Janeiro, RJ, Brazil}
\date{\today}
\maketitle

\begin{abstract}
We consider a system consisting of an
atom in the dipole approximation,  
coupled to the electromagnetic field. Using recently introduced
{\it renormalized} coordinates and {\it dressed} states, we give a non-perturbative 
solution to the atom radiation process. From this formalism a non-exponential decay 
law naturally emerges. However, at least for low frequencies and weak coupling, 
there is only a very small, practically non-detectable, 
 deviation from a purely exponential decay, in agreement 
with all previous results in the literature.

\vspace{0.34cm} \noindent
PACS Number(s):~03.65.Ca,
32.80.Pj
\end{abstract}

\section{Introduction}
Actually, apart from computer calculations, the most currently used  method to
solve coupled equations describing the physics of interacting
systems is perturbation theory. This is a common feature to
different branches of physical sciences, such as celestial
mechanics, field theory and statistical physics. The method,
originally introduced to deal with orbital problems in celestial
mechanics, has been extremely successful since then, giving
remarkably accurate results in Quantum Electrodynamics and in Weak
interactions. In high energy physics, asymptotic freedom allows to
apply Quantum Chromodynamics in its perturbative form and very
important results have been obtained in this way in the last
decades.

However, in spite of its wide applicability, there are situations
where the use of perturbation theory is not possible, as in the
low energy domain of Quantum Chromodynamics, where confinement of
quarks and gluons takes place. In this branch of physics, 
few analytical approaches in the context of Quantum field theory
are available up to the present moment (in this particular subject the 
interested reader can find a recent account in \cite{GNPLB}). There are also
other situations, in the domain of Cavity Electrodynamics and
Quantum Optics, where perturbation methods are of little
usefulness, for instance, the observation of resonant effects
associated to the coupling of atoms with strong radio-frequency
fields. The theoretical understanding of these effects on
perturbative grounds requires the calculation of very high-order
terms in perturbative series, which makes the standard Feynman
diagrams technique practically unreliable \cite{bouquinCohen}.
Methods to non-perturbatively deal with such kind of systems,
have been introduced originally in Refs. \cite{Polonsky} and
\cite{Haroche}, and since then they have been used to investigate
several situations involving the interaction of atoms and
electromagnetic fields \cite{Cohen1,Cohen2,Haroche1}.

In this paper we give a non-perturbative treatment to the system
composed of an atom (approximated by an harmonic oscillator)
interacting with the electromagnetic field. In the dipole
approximation, we reduce the problem to a system of a mechanical
oscillator (the bare atom) coupled to the infinite set of the
harmonic modes of the (bare) field. The emergence of an
exponential decay of an excited state of the oscillator and also
renormalization aspects of this problem have been thoroughly
discussed in the literature. See for instance Refs.
\cite{Thirring,ullersma,zurek,paz,caldeira,Schramm,legget,caldeira1}.
Here we use an alternative  approach, based on precisely
defined {\it renormalized} coordinates introduced and employed recently 
to describe the coupling of
a mechanical oscillator with a scalar field
in
previous papers\cite{adolfo1,adolfo2,gabriel}.  In fact,
in the quoted papers these coordinates have been denoted as
{\it dressed} coordinates but, as explained in the paragraphs
following Eq. (\ref{e22}), we presently think that the
denomination of {\it renormalized} coordinates is more appropriate for 
them. It should be
noticed that our definition of dressed states is {\it not} the
same of the collective dressed states introduced in previous works
in the literature \cite{Polonsky,Haroche}. In particular our dressed 
states are not eigenfunctions of the interacting Hamiltonian. 

To describe the
radiation process, having as initial condition that only the
mechanical oscillator (the atom) be excited, we do not proceed as
usual, considering the interaction term in the Hamiltonian written
in terms of the bare oscillator coordinates and field modes as a
perturbation, which induces transitions among the eigenstates of
the free Hamiltonian.   We introduce {\it renormalized}
coordinates for, respectively the  mechanical oscillator and the
electromagnetic field modes. In terms of these new coordinates, {\it
dressed} states can be consistently defined in section {\bf III}
and in terms of these a non-perturbative treatment of the
radiation process is possible.

Unless explicitly stated all along the paper we use natural units $c=1, \hbar=1$.

\section{The coupled system}
In this section we review the basic aspects of the method of 
reducing the system composed of a harmonic
oscillator (the atom) interacting with the electromagnetic field
to  a system of coupled harmonic oscillators. Readers familiar
with this derivation can skip the calculations 
and go directly to the comments at the end of this section.

In the Coulomb gauge,

\be
\phi=0\;,~~~\vec{\nabla}.\vec{A}=0\;,
\label{e1}
\ee
the Lagrangean for the oscillator-electromagnetic field system is
given by

\be
L=\frac{m}{2}\dot{\vec{q}}~\!^2-\frac{m\omega^2}{2}\vec{q}\!~^2+
\frac{1}{2}\int d^3\vec{x} \left[\left(\frac{\partial}{\partial
t}\vec{A}\right)^2-
\left(\vec{\nabla}\times\vec{A}\right)^2\right]+ \int
d^3\vec{x}~\!\vec{J}.\vec{A}\;, \label{e2} \ee
where the dots mean derivative with respect to time,

\be
\vec{J}=e\delta(\vec{x}-\vec{q})\dot{\vec{q}}
\label{e3}
\ee
and $e$ is the electric charge of the  oscillator. We consider the
particle-field system in a cube of side $a$, that is,
$-a/2<(x,y,z)<a/2$ and impose periodic boundary conditions for the
field: $\vec{A}(-a/2,y,z,t)=\vec{A}(a/2,y,z,t)$,
$\vec{A}(x,-a/2,z,t)=\vec{A}(x,a/2,z,t)$ and
$\vec{A}(x,y,-a/2,t)=\vec{A}(x,y,a/2,t)$. At the end we will take
the continuum limit $a\to\infty$.

Expanding the electromagnetic field in real modes we have

\be
\vec{A}(\vec{x},t)=\sum_ku_k(\vec{x})\vec{q}_k(t)\;,
\label{e4}
\ee
where the $u_k(\vec{x})$'s form a real orthonormal basis and
satisfy

\be
-\nabla^2u_k(\vec{x})=\omega_k^2\;.
\label{e5}
\ee
Replacing Eq. (\ref{e4}) in Eq. (\ref{e2}), using the identity
$\nabla.\left(\vec{F}\times \vec{G}\right)=
\vec{G}.\nabla\times\vec{F}-\vec{F}.\nabla\times\vec{G}$ and the
orthonormality relations of the $u_k(\vec{x})$'s, we find easily,

\be
L=\frac{m}{2}\dot{\vec{q}}\!~^2-\frac{m\omega^2}{2}\vec{q}~\!^2
-\frac{1}{2}\sum_{k}\left(\dot{\vec{q}}_k~\!^2-\omega_k^2\vec{q}_k~\!^2\right)
+e\sum_{k}u_k(\vec{q})\dot{\vec{q}}.\vec{q}_k\;. \label{e6}
\ee
For all dynamical purposes the above Lagrangian is equivalent to

\be
L=\frac{m}{2}\dot{\vec{q}}~\!^2-\frac{m\omega^2}{2}\vec{q}~\!^2
-\frac{1}{2}\sum_{k}\left(\dot{\vec{q}}_k~\!^2-\omega_k^2\vec{q}_k~\!^2\right)
-e\sum_{k}u_k(\vec{q}){\vec{q}}.\dot{\vec{q}}_k\;,
\label{e7}
\ee
since the difference between Eqs. (\ref{e6}) and (\ref{e7}) is  a
total time derivative.

Solving Eq. (\ref{e5}) with periodic boundary conditions we find

\be
u_k^{(1)}(\vec{x})=\sqrt{\frac{2}{V}}\sin(\vec{k}.\vec{x})\;,~~~~~
u_k^{(2)}(\vec{x})=\sqrt{\frac{2}{V}}\cos(\vec{k}.\vec{x})\;,
\label{e8}
\ee
where $V=a^3$, $\omega_k=|\vec{k}|$ and

\be
\vec{k}=\frac{2\pi}{a}(l,m,n)\;,~~~~~l,m,n=\pm 1,\pm 2,\pm ...
\ee
Notice that there are positive and negative values of $l,m,n$.
However only  one half of the total set of possible values of
$k$'s will be admitted since $\sin(\vec{k}.\vec{x})$ and
$\cos(\vec{k}.\vec{x})$ are not, respectively, linearly
independent of  $\sin(-\vec{k}.\vec{x})$ and
$\cos(-\vec{k}.\vec{x})$. Replacing Eq. (\ref{e8}) in Eq.
(\ref{e4}) and using the transversal condition $\nabla.\vec{A}=0$
we find

\be
\vec{k}.\vec{q}_k\!\!~^{(\alpha)}(t)=0\;,~~~~~\alpha=1,2\;,
\label{e9}
\ee
where $\vec{q}_k\!\!~^{(1)}(t)$ and $\vec{q}_k\!\!~^{(2)}(t)$ are
associated, respectively to $u_k^{(1)}(\vec{x})$ and
$u_k^{(2)}(\vec{x})$. From Eq. (\ref{e9}) we can write
$\vec{q}_k\!\!~^{(\alpha)}(t)=q_{k1}^{(\alpha)}\vec{\epsilon}_{k1}
+q_{k2}^{(\alpha)}\vec{\epsilon}_{k2}$, where
$\vec{\epsilon}_{k1}$ and $\vec{\epsilon}_{k2}$ are the unit
orthogonal polarization vectors. Next we consider the dipole
approximation, that is, we assume that the electromagnetic field
does not changes significantly for points around the oscillator
equilibrium position $\vec{q}=0$. This allows us to make in the
last term of Eq. (\ref{e7}) the substitution $u_k(\vec{q})\approx
u_k(0)$. From Eq. (\ref{e8}) we find $u_k^{(1)}(0)=0$, from which
we conclude that, in the dipole approximation, the modes
$\vec{q}_{k}\!\!~^{(1)}(t)$, do not interact with the oscillator.
On the other hand $u_k^{(2)}(0)=\sqrt{\frac{2}{V}}$. Also, if we
choose one of the polarization vectors to lye on the  plane
defined by $\vec{k}$ and $\vec{q}$ we can see that only the mode
corresponding to this polarization vector interacts with the
oscillator since the other mode will be orthogonal to $\vec{q}$.
If we choose $\vec{\epsilon}_{k1}$ as orthogonal to $\vec{q}$,
then the terms different from zero in the last sum of Eq.
(\ref{e7}) will contain terms proportional to
$\vec{q}.\vec{\epsilon}_{2k}\dot{q}_{2k}^{(2)}= \sin\theta_k
\dot{q}_{2k}^{(2)}q$ where $\theta_k$ is the angle between
$\vec{q}$ and $\vec{k}$. Then  Eq. (\ref{e7}) can be cast in the
form,

\be
L=\frac{m}{2}\dot{q}^2-\frac{m\omega^2}{2}q^2
-\frac{1}{2}\sum_{k}\left(\dot{q}_k^2-\omega_k^2q_k^2\right)
-e\sqrt{\frac{2}{V}}\sum_{k}\sin\theta_k\dot{q}_k q\;, \label{e10}
\ee
where we have written only the field modes that interact with the
the oscillator (without labels, since there is just only one) and
we have suppressed the vectorial symbol in the oscillator
coordinates since now only appears its modulus. From Eq.
(\ref{e10}) we obtain the Hamiltonian,

\be
H=\frac{p^2}{2m}+\frac{m\omega^2}{2}q^2+
\frac{1}{2}\sum_{k}\left(p_k^2+\omega_k^2q_k^2\right)
+e\sqrt{\frac{2}{V}}\sum_{k}\sin\theta_k p_k
q+\frac{e^2}{V}\sum_{k}\sin\theta_k q^2\;.
\label{e11}
\ee
Notice that in the above Hamiltonian the last term proportional to
$q^2$ is infinity. This infinite term  guarantees the positivity
of the Hamiltonian. In the case of an oscillator coupled to a
massless scalar field, considered in Refs.
\cite{adolfo1,adolfo2,gabriel}, this term is absent at the
beginning and is introduced afterwards by renormalizing the
oscillator frequency. Indeed, also in the present situation we can
define a  bare frequency $\omega_B$, such that,

\be
\omega_B^2=\omega^2+\frac{2e^2}{mV}\sum_{k}\sin\theta_k\;.
\label{e12}
\ee
By performing a canonical transformation,

\be
q\to\frac{q_0}{\sqrt{m}},~~~p\to\sqrt{m}p_0~~~ {\rm
and}~~q_k\to \frac{p_k}{\omega_k},~~~ p_k\to-\omega_kq_k
\label{e13}
\ee
in Eq. (\ref{e11}) we get

\be
H=\frac{p_0^2}{2}+\frac{\omega_B^2}{2}q_0^2
+\frac{1}{2}\sum_{k}\left(p_k^2+\omega_k^2q_k^2\right)
-\sum_{k}c_k q_k q\;, \label{e14}
\label{qme}
\ee
where

\be
c_k=e\sqrt{\frac{2}{mV}}\sin\theta_k\omega_k\;.
\label{e15}
\ee
The Hamiltonian given by Eq. (\ref{e14}) is almost identical to
the corresponding Hamiltonian in the scalar field case
\cite{adolfo1,adolfo2,gabriel}, with the differences that the physical 
oscillator frequency arises naturally (that is, we do not need to
renormalize the frequency oscillator), the coupling involves an
angle $\theta_k$ and the sum is in three dimensions rather than in
one. We have to mention that a Hamiltonian of the form given in
Eq. (\ref{e14}) has been used many times in the literature as a
simplified model to the quantum Brownian motion
\cite{feynman,caldeira} and also to study the decoherence problem
\cite{zurek}. However we would like to remark that in all early
treatments no use of {\it renormalized} coordinates (to be defined
below) has been done in the way we employ them. Basically we introduce 
our  renormalized coordinates in
order to guarantee the stability of the ground state of the
oscillator (the atom) in the absence of field quanta (the vacuum), 
(see Refs.
\cite{adolfo1,adolfo2} for details). If these coordinates were not
introduced, as explained below, the oscillator in its ground state
would decay to other states, in contradiction with experimental facts.

\section{Renormalized coordinates and dressed states}

We can diagonalize the Hamiltonian given by Eq. (\ref{e14}) as in
Refs. \cite{adolfo1,adolfo2,gabriel} (see also appendix of Ref.
\cite{gabrielrudnei}), defining $q_\mu=\sum_rt_\mu^rQ_r$,~
$p_\mu=\sum_r t_\mu^rP_r$,~ $\mu=(0,k)$, where $\{t_\mu^r\}$ is an
orthogonal matrix. We obtain,

\bqn
H=\frac{1}{2}\sum_r\left(P_r^2+\Omega_r^2Q_r^2\right)\;.
\label{e16}
\eqn
The matrix elements $t_\mu^r$ are given by

\be
t_0^r=\left[1+\sum_k\frac{c_k^2}{(\omega_k^2-\Omega_r^2)^2}\right]^{-\frac{1}
{2}}\;,
\label{e17}
\ee
\be
t_k^r=\frac{c_k}{(\omega_k^2-\Omega_r^2)}t_0^r
\label{e18}
\ee
and the normal frequencies $\Omega_r$, are given as solutions of
the equation

\be
\omega_B^2-\Omega_r^2=\sum_k\frac{c_k^2}{\omega_k^2-\Omega_r^2}\;.
\label{e19}
\ee
Using Eq. (\ref{e12}) and (\ref{e15}) we can write Eq. (\ref{e19})
as

\be
\omega^2-\Omega_r^2=\Omega_r^2\sum_k\frac{c_k^2}{\omega_k^2(\omega_k^2-
\Omega_r^2)}\;.
\label{e20}
\ee
In order to define the renormalized coordinates and dressed states
we start from the eigenstates of our system, represented by the
normalized eigenfunctions,

\begin{equation}
\phi_{n_{0}n_{1}n_{2}...}(Q)=\prod_{r}\left[\frac{H_{n_{r}}(\sqrt{\Omega_{r}}Q_
{r})}
{\sqrt{2^{n_r}n_{r}!}} \right]\Gamma_{0}(Q)\;,
\label{autofuncoes}
\end{equation}
where $H_{n_{r}}$ stands for the $n_{r}$-th Hermite polynomial and

\be
\Gamma_{0}(Q)\propto e^{-\frac{1}{2}\sum_r\Omega_rQ_r^2}
\label{ext1}
\ee
is the ground state eigenfunction of the system. As in Refs.
\cite{adolfo1,adolfo2,gabriel} we introduce  {\it renormalized}
coordinates $q'^{\mu}$, $\mu=\,0,\;k=1,2,...$,

\be
\sqrt{\omega_\mu}q_\mu'=\sum_rt_\mu^r\sqrt{\Omega_r}Q_r\;,
\label{e21}
\ee
where $\omega_\mu=(\omega,\{\omega_k\})$. In terms of the  renormalized
coordinates we can define our {\it dressed} states as the
complete orthonormal set of functions \cite{adolfo1},

\begin{equation}
\psi_{m_{0} m_{1}m_2...}(q^{\prime})=
\prod_{\mu}\left[\frac{H_{m_{\mu}}
(\sqrt{\omega_\mu}q^{\prime}_\mu)}
{\sqrt{2^{m_\mu}m_\mu!}}\right]\Gamma_0(q')\;,
\label{ortovestidas1}
\end{equation}
where $\Gamma_0(q')$ has the same functional form as
$\Gamma_0(Q)$, that is,

\be
\Gamma_0(q')\propto e^{-\frac{1}{2}\sum_\mu\omega_\mu
q_\mu'^2}\;.
\label{e22}
\ee
It is worthwhile to notice that our renormalized coordinates are
{\it new} objects, {\it different} from both the bare coordinates
$q$ and the normal coordinates $Q$, and have been used for the
first time in Refs. \cite{adolfo1,adolfo2,gabriel}. In particular,
the renormalized coordinates and dressed states although being
collective objects, should {\it not} be confused with the normal
coordinates $Q$, and the eigenstates (\ref{autofuncoes}).
Differently from  the eigenstates (\ref{autofuncoes}) which are
stable, our dressed states are all unstable, except the ground 
dressed state obtained by setting $\{m_\mu=0\}$ in Eq.
(\ref{ortovestidas1}).

The idea, introduced in Ref. \cite{adolfo1}, (for a recent clear 
explanation see Ref. \cite{yony}) 
is that the dressed states given by
Eq. (\ref{ortovestidas1}) are supposed to be the physically
meaningful states, instead of the ones written in terms of the
{\it bare}\ coordinates $q_\mu$ that appear in the original
Lagrangean. This can be seen as an analogous of the wave function
renormalization in quantum field theory, which justifies the
denomination of {\it renormalized} to the new coordinates $q'$.
Thus, the dressed state given by Eq. (\ref{ortovestidas1})
describes the atom in its $m_0$-th excited level and $m_k$ field
quanta (photons) of frequencies $\omega_k$. It should be noticed
that the introduction of the renormalized coordinates guarantees
the stability of the dressed vacuum state, since by definition it
is identical to  ground state of the system. The fact that the
definition given by Eq. (\ref{e21}) assures this requirement, can
be easily seen replacing Eq. (\ref{e21}) in Eq. (\ref{e22}). We
obtain $\Gamma_0(q')\propto\Gamma_0(Q)$, which shows that the
dressed vacuum state given by Eq. (\ref{e22}) is the same ground state
of the interacting Hamiltonian given by Eq. (\ref{e16}).

The necessity of introducing renormalized coordinates can be
understood by considering what would happen if we write Eq.
(\ref{ortovestidas1}) in terms of the bare coordinates $q_\mu$. In
the absence of interaction, the bare states are stable since they are
eigenfunctions of the Hamiltonian obtained from Eq. (\ref{e14}) by
setting $c_k=0$. But when we consider the interacting term they
become all unstable. With the excited states there is no problem
since we known from experiment that this is indeed  the case. On the other
hand, we know from experiment that the atom in its ground state is
stable, in contradiction with what our simplified model for the 
system describes 
in terms of the bare 
coordinates. Obviously our simplified model is only approximate, the
correct theory that describes the system atom-electromagnetic field is Quantum
Electrodynamics ($QED$), but a rigorous, non- 
perturbative description using $QED$ would be extremely  
complicated. So, if we wish to have a non-perturbative approach in terms of 
our simplified model we have
to modify something in order to remedy the above mentioned problem. 
The solution is just the introduction of the renormalized
coordinates $q_\mu'$, as the physically meaningful ones. 
On the
other hand if, in terms of  creation and annihilation 
operators,
we use the so called rotating wave approximation
\cite{milonni} for the interacting term in Eq. (\ref{e14}) 
(the Lee-Friedrichs model),  we would
remain, neglecting part of the interaction Hamiltonian,  
only with terms involving $\hat{a}_0^\dag\hat{a}_k$ and
$\hat{a}_k^\dag\hat{a}_0$ [this is not the case for the whole Hamiltonian 
(\ref{e14}) which we consider here].
 In this situation there is no problem with
the stability of the atom ground state described in terms of
the bare coordinates $q_\mu$, since in this case the action of the
interacting term on the bare oscillator ground state (obtained
from Eq. (\ref{ortovestidas1}) setting $q_\mu'=q_\mu$ and
$m_\mu=0$), gives zero. That is, the free oscillator ground state
still remain as the ground state of the interacting system. This
is the reason why in early treatments (see for example
\cite{prigogine} and references therein) of the simplified model presented here, 
 there has been no
need to introduce renormalized coordinates. 

We can write the Hamiltonian (\ref{qme}) in terms of dressed coordinates.
Replacing Eq. (\ref{e21}) in Eq. (\ref{qme}) we get the renormalized Hamiltonian,
\bqn
H&=&\frac{1}{2}\left(\alpha_{00}(p_0')^2+\beta_{00}\omega^2(q_0')^2\right)+\frac{1}{2}
\sum_{k}\left(\alpha_{kk}(p_k')^2+\beta_{kk}\omega_k^2(q_k')^2\right)
+\sum_{k}\left(\alpha_{0k}p_0'p_k'+\beta_{0k}\omega_0\omega_kq_0'q_k'\right)\nonumber\\
& &+\sum_{k_1>k_2}\left(\alpha_{k_1k_2}p_{k_1}'p_{k_2}'
+\beta_{k_1k_2}\omega_{k_1}\omega_{k_2}q_{k_1}'q_{k_2}'\right)
\;,
\label{qme1}
\eqn
where
\be
\alpha_{\mu\nu}=\sqrt{\omega_\mu\omega_\nu}\sum_{r}\frac{t_\mu^rt_\nu^r}{\Omega_r}
\label{qme2}
\ee
and
\be
\beta_{\mu\nu}=\frac{1}{\sqrt{\omega_\mu\omega_\nu}}
\sum_{r}\Omega_rt_\mu^rt_\nu^r\;.
\label{qme3}
\ee
Expression (\ref{qme1}) seems
complicated and without clear physical significance. 
In particular, it appears that the dressed
atom and the dressed field modes are in the same foot of equality, that is,
the dressed atom, described by $q_0'$ and $p_0'$, appears to
play no central role as in the original Hamiltonian. 
However this is not
the case. Since the expressions for $t_0^r$ and $t_k^r$, given by Eqs. 
(\ref{e17})-(\ref{e18}), are very different, the renormalized coordinate $q_0'$ and
momentum $p_0'$ still play a central role, because they are multiplied respectively
by $\alpha_{00}$ and $\beta_{00}$ that contain
terms of the type $(t_0^r)^2$, see Eqs. (\ref{qme2})-(\ref{qme3}).  On the other
hand the dressed coordinates $q_k'$ and momenta $p_k'$
are multiplied by $\alpha_{kk}$ and $\beta_{kk}$ than contain terms of the type
$t_{k}^rt_{k}^r$.
The interacting terms in Eq. (\ref{qme1}) involve both dressed coordinates
and momenta different from the original Hamiltonian, that involve only 
{\it explicit}
interaction among the coordinates $q_0$ and $q_k$. The last term in the first
line of Eq. (\ref{qme1}) expresses {\it explicit} interaction between the dressed
atom and the dressed field modes. The second line in Eq. (\ref{qme1})
expresses the {\it explicit} interaction among the dressed field modes.
However the couplings of these two type of interactions are very different,
the dressed atom-photon and dressed photon-photon interactions 
contain, respectively, terms like $t_0^rt_k^r$ and $t_{k_1}^rt_{k_2}^r$ that
are very different. 

We can show that the Hamiltonian written in dressed coordinates,
given by Eq. (\ref{qme1}), 
approximately decouples into the dressed atom and the dressed field modes.
That is, we show that in Eq. (\ref{qme1}),
$\alpha_{\mu\mu}\gg\alpha_{\mu\nu}|_{\mu\neq\nu}$ and 
$\beta_{\mu\mu}\gg\beta_{\mu\nu}|_{\mu\neq\nu}$. For this end note that
for $c_k=0$ (or equivalently for $e=0$, where $e$ is the electric charge)
$q_\mu=q_\mu'$. In this case, $\alpha_{\mu\mu}=\beta_{\mu\mu}=1$ and 
$\alpha_{\mu\nu}|_{\mu\neq\nu}=\beta_{\mu\nu}|_{\mu\neq\nu}=0$.
Then, because $e\ll 1$,  we must have
$\alpha_{\mu\mu}=1+{\cal O}(e)$, and $\beta_{\mu\mu}=1+{\cal O}(e)$ 
whereas $\alpha_{\mu\nu}|_{\mu\neq\nu}={\cal O}(e)$ and 
$\beta_{\mu\nu}|_{\mu\neq\nu}={\cal O}(e)$,
showing that $\alpha_{\mu\mu}\gg\alpha_{\mu\neq\nu}$ and 
$\beta_{\mu\mu}\gg\beta_{\mu\neq\nu}$. This shows that the
introduction of the renormalized coordinates allows, approximately, 
to divide the system into the dressed atom and the dressed field modes.

The decomposition given in Eq. (\ref{qme1}) shows that the renormalized 
Hamiltonian takes into account {\it explicitly}
some of the {\it implicit} interactions that are preset in the original Hamiltonian.
For example, a photon-photon interaction term, of the type given by the second
line of Eq. (\ref{qme1}) can be obtained from the interaction term $q_0q_k$,
that appears in the original Hamiltonian,  
at second order in perturbation theory. Also we have to note that in 
Eq. (\ref{qme1}) the "uncoupled" terms, are not identical to the
uncoupled ones in Eq. (\ref{qme}), the differences are just self-interacting
terms. From these discussions we can see that our dressed states, 
defined by Eq. (\ref{ortovestidas1}), are not free, they are collective states, 
that take into account the interactions. The dressed atom, for example is the
bare atom  "surrounded" by bare (virtual)
photons.  The same picture applies to the dressed 
photons. That these claims are correct can be seen naturally, writing any dressed
state given by Eq. (\ref{ortovestidas1}) in the basis of bare states, the
bare states being identical to the states given by Eq. (\ref{ortovestidas1}) with
the dressed coordinates $q_\mu'$ replaced by the bare ones $q_\mu$.

With the formalism of dressed coordinates and dressed states we
can compute the probability that the atom (understood as the
dressed mechanical oscillator $q'_{0}$) prepared at $t=0$ in its
first excited state, still remain at the time $t$ in the same
state, or from it the probability of decay of the atom from its
first excited state. For this end we consider the initial dressed
state $\psi_{100...}(q^{\prime})$, in which only the dressed atom
is in the first excited level, all the others oscillators being in
the ground state. Then it is shown in Ref. \cite{adolfo1} the
following expression for its time evolution,

\begin{equation}
\psi_{100...}(q^{\prime},t)=\sum_{\mu}f_{0
\mu}(t)\psi_{0...01_\mu0...}(q^{\prime}) \label{ortovestidas5}
\end{equation}
where
\begin{equation}
f_{0\mu}(t)=\sum_{s}t_{0}^{s}t_{\mu}^{s}e^{-i\Omega_{s}t}.
\label{added1}
\end{equation}
From Eq. (\ref{ortovestidas5}) we see that the initially excited
dressed atom naturally distributes its energy among itself and all
other dressed oscillators (the dressed field quanta) as time goes
on, with well defined probability amplitudes given by the
quantities $f_{0 \mu}(t)$ in Eq. (\ref{ortovestidas5}).
From Eq. (\ref{ortovestidas5}) we obtain for the probability
amplitude that the dressed mechanical oscillator (the atom) remain in its
first excited state the following expression,

\be
f_{00}(t)=\sum_r (t_0^r)^2 e^{-i\Omega_r t}\;.
\label{e23}
\ee

As already mentioned above, we do not make explicit use of the
concepts of interacting bare oscillator and field, described by
the bare coordinates $q_{0}$ and $\{q_{k}\}$. We have introduced
renormalized coordinates $q^{\prime}_{0}$ and $\{q^{\prime}_{k}\}$
for, respectively the dressed atom and the dressed field modes. In
terms of these new coordinates dressed  states can be rigorously
defined. A non-perturbative approach to the radiation process  is
possible in terms of the {\it dressed} states for the system
contained in a cavity of arbitrary size, including microcavities
\cite{adolfo2}. Also free space emission starting from periodic
boundary conditions, which we examine in this paper can be treated
using our dressed states. This unified treatment is not possible
in the conventional framework.  Our dressed states
Eq. (\ref{ortovestidas1}) are {\it collective} but {\it non
stable} states, linear combinations of the eigenstates
(\ref{autofuncoes}) defined in terms of the normal modes. The
coefficients of these combinations  can be calculated and explicit
formulas for these coefficients for the interesting states are
given in Refs. \cite{adolfo1,adolfo2}. Moreover, as noticed above,
they  have the very interesting property, used in Refs.
\cite{adolfo1,adolfo2,gabriel} for the case of a scalar field, of
distributing the energy initially in a particular dressed state,
among itself and all other dressed states with precise and
calculable time-dependent probability amplitudes. We {\it choose}
these dressed states as physically meaningful to the study of the
radiation process.

\section{The radiation process}
Now, to compute Eq. (\ref{e23}) in the continuum limit
$a\to\infty$, let us define the complex valued function,

\be
W(z)=z^2-\omega^2+z^2\sum_k\frac{c_k^2}{\omega_k^2(\omega_k^2-z^2)}\;.
\label{e24}
\ee
Notice, from Eqs. (\ref{e20}) and (\ref{e24}), that the
$\Omega_r$'s are given by the real roots of $W(z)$, and that in
the continuum limit these real roots will extend over all the
positive real axis in the complex $z$-plane. Deriving Eq.
(\ref{e24}) with respect to $z$ and comparing with Eq. (\ref{e17})
we find,

\be
(t_0^r)^2=\left.\frac{2z}{W'(z)}\right|_{z=\Omega_r}\;.
\label{e25}
\ee
Replacing Eq. (\ref{e25}) in Eq. (\ref{e23})and using  Cauchy
theorem we get

\be
f_{00}(t)=\frac{1}{\pi i}\oint_{\cal C} dz \frac{z}{W(z)}\;,
\label{e26}
\ee
where ${\cal C}$ is a counterclockwise contour in the $z$-plane
that encircles the real positive axis. The integral in Eq.
(\ref{e26}) can be computed choosing a contour ${\cal C}$ that
lies just below and above  the real positive axis. Below the real
positive axis we have $z=\Omega-i\epsilon$ and above
$z=\Omega+i\epsilon$. Then we have

\be
f_{00}(t)=\frac{1}{i\pi}\int_0^\infty
d\Omega\left[\frac{(\Omega-i\epsilon)
e^{-(i\Omega+\epsilon)t}}{W(\Omega-i\epsilon)}-
\frac{(\Omega+i\epsilon)
e^{-(i\Omega-\epsilon)t}}{W(\Omega+i\epsilon)}\right]\;.
\label{e27}
\ee
The next step is to compute $W(\Omega+i\epsilon)$ and
$W(\Omega-i\epsilon)$. Using Eq. (\ref{e15}) in Eq. (\ref{e24}) we
get in the continuum limit

\bqn
W(z)&=&z^2-\omega^2+\frac{e^2z^2}{4m}\int_0^{2\pi}d\phi\int_0^{\pi/2}d\theta_k
\int_0^\infty dk\frac{k^2\sin\theta_k^3}{k^2-z^2}\nonumber\\
&=&z^2-\omega^2+\frac{e^2z^2}{3\pi^2m}\int_0^\infty
dk\frac{k^2}{k^2-z^2}\;,
\label{e28}
\eqn
where  we integrate over the $\theta_k$ angle only up to $\pi/2$
instead to $\pi$, since as we remarked above, only one half
of the modes are linearly independent. The integral over $k$ in
Eq. (\ref{e28}) is linearly divergent. Then in order to make
meaningful the above equation we will need to regularize it. We do
this by analytic regularization, that is, we introduce the
integral

\be
I(\alpha)=\mu^{\alpha-1}\int_0^\infty
dk\frac{k^2}{(k^2-z^2)^\alpha}\;, \label{e29}
\ee
which is well defined for $\alpha>2$~;  then we can give a meaning
to the integral in Eq. (\ref{e28}) by analytically extending the
integral given by Eq. (\ref{e29}) to $\alpha=1$. The parameter
$\mu$ in Eq. (\ref{e29}) is introduced in order to maintain the
correct dimension of the original integral. Using Cauchy theorem
we obtain for $I(\alpha)$,

\bqn
I(\alpha)&=&i\pi\mu^{\alpha-1}(\alpha-1)!\left.\frac{\partial^{\alpha-1}}
{\partial w^{\alpha-1}}\frac{w^2}{(w+z)^\alpha}\right|_{w=z}\;,~~~Im(z)>0\;,\\
I(\alpha)&=&-i\pi\mu^{\alpha-1}(\alpha-1)!\left.\frac{\partial^{\alpha-1}}
{\partial
w^{\alpha-1}}\frac{w^2}{(w+z)^\alpha}\right|_{w=z}\;,~~~Im(z)<0\;.
\label{e30}
\eqn
Then by analytic extension we find $I(1)=i\pi z/2$ for $Im(z)>0$
and $I(1)=-i\pi z/2$ for $Im(z)<0$. Replacing these values in the
integral of Eq. (\ref{e28}) we find

\bqn
W(\Omega+i\epsilon)&=&\Omega^2-\omega^2+i\frac{e^2\Omega^3}{6\pi
m}+{\cal O}(\epsilon)\;,\\
W(\Omega-i\epsilon)&=&\Omega^2-\omega^2-i\frac{e^2\Omega^3}{6\pi
m}+ {\cal O}(\epsilon).
\eqn

Substituting the above expressions in Eq. (\ref{e27}) we get in
the $\epsilon\to 0^+$ limit,

\be
f_{00}(t)=\frac{2\tau}{\pi}\int_0^\infty d\Omega
\frac{\Omega^4 e^{-i\Omega t}}
{(\Omega^2-\omega^2)^2+\tau^2\Omega^6}\;,
\label{e31}
\ee

where we have introduced the time dimensional parameter
$\tau=e^2/(6\pi m)$. The integral given by Eq. (\ref{e31}) can not
be evaluated analytically. We can check the validity of Eq.
(\ref{e31}) taking $t=0$ and noticing that for $\tau\omega\ll 1$
the integrand is sharply peaked around $\Omega=\omega$. Then we
obtain from Eq. (\ref{e31}),

\bqn
f_{00}(0)&=&\frac{\tau\omega^3}{\pi}\int_{-\infty}^{\infty}dx
\frac{1}{x^2+\tau^2\omega^6}\nonumber\\
&=&1\;, \label{e32}
\eqn
as expected from the definition given by Eq. (\ref{e23}). One can
check the above result also for any value of $\tau$ using Cauchy
theorem for $t=0$ in Eq. (\ref{e31}). We see from Eq. (\ref{e31})
that the probability $|f_{00}(t)|^2$ will not be a pure exponential.
However, at least for low frequencies as we will see below, there
will be only a small deviation from an exponential law.

\begin{figure}[c]
\epsfysize=6cm 
\centerline{\epsfbox{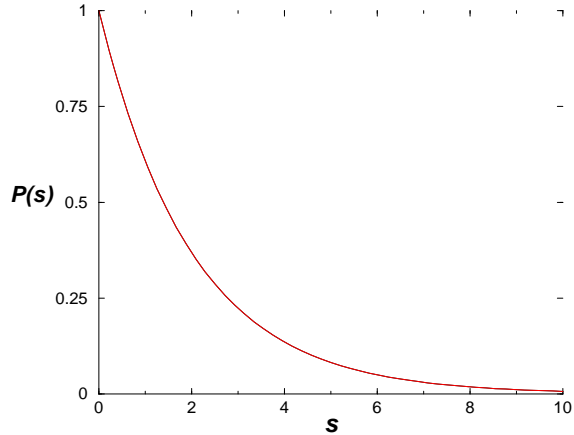}}
\caption{Plot on the vertical axis, of the probability that the
oscillator remain excited at time $t$ in units $\omega t=s$.}
\label{eletrofig}
\end{figure}

Considering $t>0$ we can integrate Eq. (\ref{e31}) numerically and
from the result compute the probability $P(t)=|f_{00}(t)| ^2$ that
the atom still remain excited at time $t$. For instance,
taking $\tau \omega=0.05$ and using for $e$ and $m$ respectively
the proton charge and the mass of the hydrogen atom, we get
$\omega \approx 5.9\times 10^{10}/s$.  For these values we obtain  the plot 
showed in Fig. \ref{eletrofig} where we can see an almost exponential behavior
for the decaying process.

\section{Concluding Remarks}
In this paper we have analysed a  simplified version of an
atom-electromagnetic field system and we have tried to give the
more exact and rigorous treatment we could to the problem. We have
adopted a general physicist's point of view, in the sense that we
have renounced to approach very closely to the real behavior of a
complicated non-linear system. As a counterpart, an exact solution
has been possible. We have used  a formalism ({\it renormalized}
coordinates and  {\it dressed} states) that allows a non-perturbative approach
to the time evolution of the system. In particular, we have
applied this formalism to study non perturbatively the time
evolution of the excited atom. In the free space  limit,  we get
the result that the renormalized coordinate approach describing
the atom modified by the presence of the field in an indissoluble
way, gives an exact result for emission in free space,
generalizing the well known exponential decay law.

{\bf Acknowledgements}\\
GFH is supported by FAPESP grant 02/09951-3 and APCM is partially
supported by CNPq (Brazilian National Research Council).


\begin{thebibliography}{99}


\bibitem{GNPLB} A. P. C. Malbouisson, A. E. Santana, J. M. C. Malbouisson
and J. C. da Silva Phys. Lett. B{\bf 583}, 373 (2004).

\bibitem{bouquinCohen} C. Cohen-Tannoudji, {\it Atoms in Electromagnetic 
Fields},
(World Scientific publishing Co. 1994).

\bibitem{Polonsky} N. Polonsky, Doctoral thesis, Ecole Normale Sup{\'e}rieure,
Paris (1964).

\bibitem{Haroche} S. Haroche, Doctoral thesis, Ecole Normale Sup{\'e}rieure,
Paris (1964).

\bibitem{Cohen1}  C. Audoin, C. Cohen-Tannoudji, S. Haroche and J. P. Schermann,
Phys. Rev. Lett. {\bf 24} 861 (1970).

\bibitem{Cohen2}  C. Cohen-Tannoudji and S. Haroche, Phys. Rev. Lett. {\bf 24}
974 (1970)

\bibitem{Haroche1}  S. Haroche and  S. Liberman, {\it Frontiers in Laser
Spectroscopy}, eds. R. Balian (North Holland, 1977).

\bibitem{Thirring} W. Thirring and F. Schwabl, Ergeb. Exakt. Naturw. {\bf 36},
219 (1964).

\bibitem{ullersma} P. Ullersma, Physica {\bf 32}, 56 (1966); Physica {\bf 32},
74 (1966); Physica {\bf 32}, 90 (1966).

\bibitem{zurek} W. G. Unruh, W. H. Zurek, Phys. Rev. D, {\bf 40}, 1071
(1989).

\bibitem{paz} B. L. Hu, J. P. Paz and Y. Zhang, Phys. Rev. D{\bf 45},
2843 (1992).

\bibitem{caldeira} A. O. Caldeira and A. J. Legget, Ann. Phys. (N.Y) {\bf 149},
374 (1983).

\bibitem{Schramm} H. Grabert, P. Schramm and G.-L. Ingold, Phys. Rep. {\bf 168},
115 (1988).

\bibitem{legget} A. J. Legget et al., Rev. Mod. Phys. {\bf 59}, 1
(1987).

\bibitem{caldeira1} A. O. Caldeira and A. J. Legget, Physica A {\bf 121}, 587 
(1983).

\bibitem{adolfo1} N. P. Andion, A. P. C. Malbouisson and A. Mattos Neto,
J. Phys.  A{\bf 34}, 3735 (2001).

\bibitem{adolfo2} G. Flores-Hidalgo, A.P.C. Malbouisson and Y.W. Milla,
Phys. Rev. A {\bf 65}, 063414 (2002).

\bibitem{gabriel} G. Flores-Hidalgo and A. P. C. Malbouisson, Phys. Rev. 
A{\bf 66}, 042118 (2002).

\bibitem{feynman} R. P. Feynman and F. L. Vernon Jr., Ann. Phys. {\bf 24}, 118 
(1963);
Ann. Phys. {\bf 281}, 547 (2000).

\bibitem{gabrielrudnei} G. Flores-Hidalgo and R. O. Ramos,  Physica A{\bf 326},
159 (2003), hep-th/0206022.

\bibitem{yony} G. Flores-Hidalgo and Y. W. Milla, {\it Dressed (renormalized)
coordinates in a nonlinear system}, arXiv: physics/0410238.

\bibitem{milonni} P. W. Milonni, {\it The quantum vacuum},
(Academic Press, Inc., New York, 1994).

\bibitem{prigogine} S. Kim and G. Ordonez, Phys. Rev. E{\bf 67},
056117 (2003).

\end{thebibliography}
\end{document}